\pdfoutput=1
\documentclass[preprint,12pt]{elsarticle}

\usepackage[letterpaper, top=1in, bottom=1in, left=1.25in, right=1.25in]{geometry}
\usepackage{graphicx}
\usepackage{hyperref}
\usepackage{booktabs}
\usepackage{multirow}
\usepackage{bm}
\usepackage{xcolor}
\usepackage{amsmath,amssymb}
\usepackage{lineno}
\usepackage{float}
\usepackage{placeins}

\newcommand\norm[1]{\left\lVert#1\right\rVert}

\journal{Physics of Plasmas}

\begin{document}

\begin{frontmatter}

\title{A Pre-trained Variational Autoencoder for Gyrokinetic Plasma Turbulence Surrogate Modeling}

\author[ornl]{Minglei Yang\corref{cor1}}
\ead{yangm@ornl.gov}
\cortext[cor1]{Corresponding author}
\author[gvsu]{Marshall Nicholson}
\author[ut]{Diego Del-Castillo-Negrete}
\author[ut]{David Hatch}
\author[ornl]{Guannan Zhang}

\affiliation[ornl]{organization={Oak Ridge National Laboratory}, city={Oak Ridge}, state={TN}, postcode={37831}, country={USA}}
\affiliation[gvsu]{organization={Grand Valley State University}, city={Allendale}, state={MI}, postcode={49401}, country={USA}}
\affiliation[ut]{organization={University of Texas at Austin}, city={Austin}, state={TX}, postcode={78712}, country={USA}}

\begin{abstract}
Machine learning surrogate models offer a promising path toward accelerating plasma turbulence simulations. We present PreVAE-Turb, a surrogate modeling framework that leverages pre-trained variational autoencoders (VAEs) from the Stable Diffusion image generation model for efficient spatial compression of turbulence fields. The pre-trained VAE is fine-tuned on turbulence data using a physics-informed loss function that includes a spectral loss operating in Fourier space to enforce spectral accuracy across scales. The VAE is combined with convolutional long short-term memory (ConvLSTM) networks to learn temporal dynamics in latent space, with a manifold consistency error metric that monitors encode--decode consistency during autoregressive rollouts. We validate the framework on two-dimensional Hasegawa-Wakatani drift-wave turbulence and extend it to gyrokinetic turbulence from the GENE code, where a four-channel adaptation simultaneously predicts electrostatic potential, density, and parallel/perpendicular temperature fluctuations without requiring architecture redesign. Once trained, inference generates thousands of time steps in seconds on a single GPU, providing substantial computational acceleration compared to direct numerical simulation. The pre-trained approach offers a transferable methodology broadly applicable to various turbulence simulation codes.
\end{abstract}

\tnotetext[fn1]{{\bf Notice}:  This manuscript has been authored by UT-Battelle, LLC, under contract DE-AC05-00OR22725 with the US Department of Energy (DOE). The US government retains and the publisher, by accepting the article for publication, acknowledges that the US government retains a nonexclusive, paid-up, irrevocable, worldwide license to publish or reproduce the published form of this manuscript, or allow others to do so, for US government purposes. DOE will provide public access to these results of federally sponsored research in accordance with the DOE Public Access Plan.}
\end{frontmatter}

\section{Introduction}
\label{sec:intro}

Turbulence-driven transport is a critical phenomenon that governs confinement performance in magnetically confined fusion plasmas.
In tokamaks and stellarators, microinstabilities drive turbulent fluctuations that transport particles, momentum, and heat across magnetic field lines, with the resulting anomalous transport often exceeding classical and neoclassical predictions by orders of magnitude.
The inherently nonlinear nature of plasma turbulence---where fluctuations interact across disparate spatial and temporal scales---makes this a central challenge for magnetic confinement fusion.
Understanding and predicting this turbulent transport is therefore essential for optimizing plasma performance and designing future fusion reactors.

First-principles simulation of plasma turbulence requires solving the nonlinear gyrokinetic equations, which describe the evolution of the gyro-center distribution function in five-dimensional phase space~\cite{krommes2012gyrokinetic, brizard2007foundations}. Several sophisticated codes have been developed for this purpose, including GENE~\cite{jenko2000electron, gorler2011gene}, GKW~\cite{peeters2009nonlinear},
CGYRO~\cite{candy2016high},
and others~\cite{kotschenreuther1995comparison, candy2003eulerian}. These codes are computationally intensive due to the high dimensionality of the phase space, the need for fine spatial and temporal resolution to capture multiscale dynamics, and the evaluation of multiple coupled fields (e.g., electrostatic potential and magnetic vector potential) for each particle species---a single ion-scale nonlinear simulation can require hundreds of GPU-hours~\cite{ashourvan2026gkfieldflow}.
This computational burden severely limits their use in scenarios requiring many evaluations, such as uncertainty quantification, optimization, and real-time control applications. Reduced models such as QuaLiKiz~\cite{bourdelle2015core}
and TGLF~\cite{staebler2007theory}
offer faster alternatives through quasilinear approximations, but sacrifice accuracy in capturing nonlinear saturation physics.

Machine learning has emerged as a powerful tool across the fusion research community. Applications include disruption prediction using deep neural networks~\cite{kates2019predicting}, real-time plasma control via reinforcement learning~\cite{degrave2022magnetic}, and automated analysis of diagnostic data~\cite{ferreira2019deep}. For turbulent transport prediction, neural network surrogates have been developed to emulate quasilinear models: QLKNN provides rapid flux predictions by learning QuaLiKiz outputs~\cite{van2020fast}, and similar approaches have been applied to accelerate TGLF-based transport modeling~\cite{cao2025tglf}. These transport surrogates predict scalar flux quantities but do not resolve the underlying turbulent field structures. Other machine learning applications target edge plasma modeling~\cite{csala2025autoregressive, zhu2022data, leppink2025optimization}, predicting transport-scale dynamics in the scrape-off layer rather than core turbulence fluctuations.
More recently, deep learning approaches have been developed to predict the full spatio-temporal evolution of turbulent fields. The GAIT framework~\cite{clavier2025generative} pioneered the VAE-RNN paradigm for 2D Hasegawa-Wakatani turbulence, training a custom encoder-decoder from scratch and establishing comprehensive statistical validation metrics. GKFieldFlow~\cite{ashourvan2026gkfieldflow} extended field-resolving surrogates to 3D gyrokinetic turbulence from CGYRO using U-Net architectures with temporal convolutional networks, focusing on single-field prediction. GyroSwin~\cite{paischer2025gyroswin} proposed a foundation model approach with approximately 1 billion parameters for the full 5D distribution function, demonstrating impressive capability but requiring substantial computational resources.

In this work, we present PreVAE-Turb, a surrogate modeling framework for plasma turbulence that leverages pre-trained variational autoencoders (VAEs) from the Stable Diffusion image generation model~\cite{rombach2022high}. Our approach builds upon the generative framework introduced by GAIT~\cite{clavier2025generative}, which demonstrated the effectiveness of combining VAEs with recurrent neural networks for Hasegawa-Wakatani turbulence~\cite{hasegawa1983plasma}. The main contributions of this work are:
1) We utilize the AutoencoderKL architecture from Stable Diffusion v1.5~\cite{von-platen-etal-2022-diffusers}, pre-trained on natural images, rather than training a custom encoder-decoder from scratch. This transfer learning strategy provides robust feature extraction for complex spatial structures, reduces the need to design a spatial encoder--decoder specifically for each turbulence dataset.
2) We introduce a physics-informed spectral loss that operates in Fourier space to enforce spectral accuracy across all scales. Unlike pixel-space losses that do not constrain the frequency content, this Fourier-domain loss ensures accurate reproduction of the energy distribution across scales---a critical requirement for turbulence surrogates that standard reconstruction losses do not address.
3) We employ convolutional LSTM (ConvLSTM) layers that preserve spatial structure in the latent space, enabling the model to capture spatially correlated temporal evolution patterns.
4) We introduce a manifold consistency error metric that monitors encode--decode consistency during long autoregressive rollouts.
5) We extend the framework to simultaneously predict multiple coupled fields (density $n$ and potential $\phi$), and demonstrate that the model accurately reproduces the cross-correlation structure between fields, which is essential for capturing turbulent transport physics.
6) We demonstrate the framework on gyrokinetic turbulence data from the GENE code~\cite{jenko2000electron,gorler2011gene}, accommodating four physical channels (electrostatic potential, density, and parallel/perpendicular temperatures) through a modified input layer while preserving the pre-trained weights.

The remainder of this paper is organized as follows. Section~\ref{sec:method} describes the methodology, including the pre-trained VAE architecture, LSTM temporal model, and training procedures. Section~\ref{sec:hw_validation} presents validation results on the Hasegawa-Wakatani model, while Section~\ref{sec:gyrokinetic} demonstrates the extension to GENE gyrokinetic simulations with detailed comparisons of spectral properties, statistical metrics, and computational performance. Finally, Section~\ref{sec:conclusion} concludes the paper and discusses future directions.

\section{Methodology: PreVAE-Turb Framework}
\label{sec:method}

We present the PreVAE-Turb framework, a modular surrogate modeling approach for multi-channel turbulence simulations. The framework is inspired by GAIT~\cite{clavier2025generative, clavier2025generativePRE}, which established the VAE--RNN paradigm for field-resolving HW turbulence, but differs in the choice of spatial representation and latent-time evolution. Specifically, we replace a from-scratch, case-specific encoder--decoder with a \emph{foundation-model VAE backbone} (AutoencoderKL from Stable Diffusion) that is fine-tuned on turbulence data, and we couple it to a latent-dynamics model for temporal prediction. This design reduces architecture engineering while enabling (i) high-fidelity compression of multi-scale fields, and (ii) a straightforward extension to multi-field inputs via minimal channel adaptation without redesigning the core backbone. In this section we describe the methodology using HW density and potential profiles as an illustrative example; the multi-field gyrokinetic application to GENE is presented in Section~\ref{sec:gyrokinetic}.

The {Hasegawa-Wakatani model} describes 2D drift-wave turbulence in magnetized plasma through coupled equations for density $n$ and vorticity $\Omega = \nabla^2 \phi$:
\begin{align}
\partial_t n &= c_1 ( \phi - n ) - [ \phi, n ] - \kappa_n \partial_y \phi - \nu (-\nabla^2)^N n, \nonumber \\
\partial_t \Omega &= c_1 ( \phi - n ) - [ \phi, \Omega ] - \nu (-\nabla^2)^N \Omega, \label{eq:hw}
\end{align}
where $[\cdot,\cdot]$ denotes the Poisson bracket. We use parameters $c_1 = 1$, $k_0 = 0.15$, $N=3$, $\nu = 5\times 10^{-8}$, on a domain $(2\pi/k_0)^2$ discretized on a $512^2$ grid and downsampled to $64^2$ for training. The dataset contains $N_s = 8000$ snapshots with time step $\Delta t = 0.025$. 

A typical turbulence simulation produces time series of multi-field snapshots, where each snapshot contains $C$ physical fields (e.g., density, potential, temperature) on a spatial grid of resolution $N_x \times N_y$ (or $N_x \times N_y \times N_z$ in 3D). Direct end-to-end learning of the mapping from past states to future states in this high-dimensional space is computationally prohibitive and often leads to poor generalization. Moreover, treating each physical field independently would neglect the important cross-field correlations that govern turbulent transport---for instance, the particle flux depends on correlations between density and potential fluctuations.

\begin{figure*}[!htbp]
    \centering
    \includegraphics[width=\textwidth]{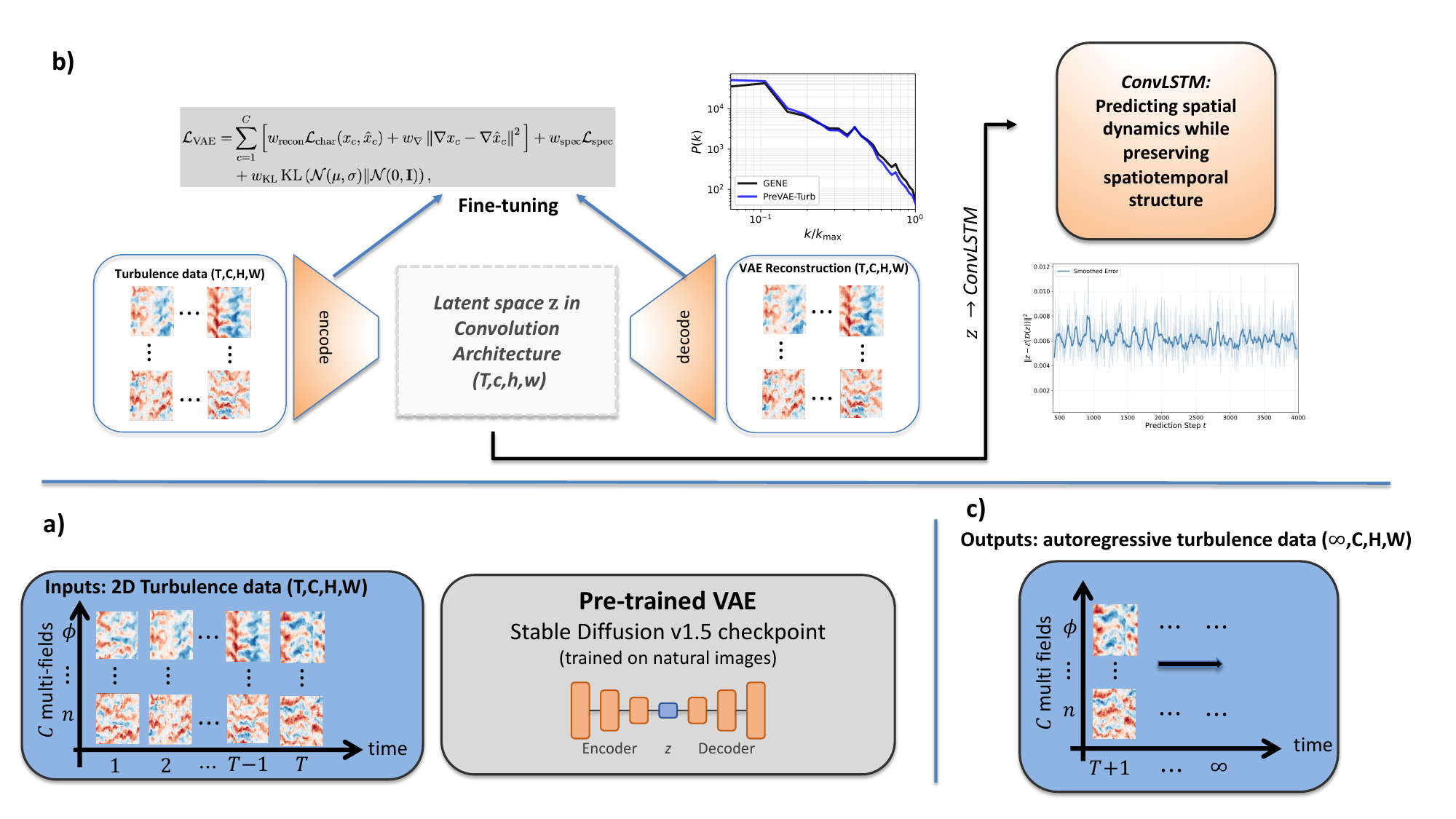}
    \caption{Overview of the PreVAE-Turb framework. \textbf{(a)} Input: multi-field 2D turbulence snapshots; the diagram labels the spatial dimensions as $H=N_x$ and $W=N_y$. The VAE backbone is initialized from Stable Diffusion v1.5~\cite{von-platen-etal-2022-diffusers}, pre-trained on natural images. \textbf{(b)} Two-stage architecture: Stage 1 fine-tunes the pre-trained VAE on turbulence data using a physics-informed loss combining reconstruction, gradient, spectral (upper right), and KL divergence terms. Stage 2 trains a ConvLSTM to predict temporal dynamics in latent space while preserving spatial correlations. The manifold consistency error (lower right) monitors encode--decode consistency of predicted latents. \textbf{(c)} Output: autoregressive prediction generates turbulence fields beyond the training horizon.}
    \label{fig:architecture}
\end{figure*}

To address these challenges, we adopt a two-stage architecture that decouples spatial compression from temporal dynamics learning, enabling each component to be optimized for its specific task while maintaining the ability to capture multi-field correlations. Figure~\ref{fig:architecture} provides an overview of the PreVAE-Turb framework.

\subsection{Two-Stage Architecture Motivation}

The two-stage approach separates the surrogate modeling problem into complementary sub-tasks:

\textbf{Stage 1: Spatial Compression via VAE.}
The first stage employs a variational autoencoder~\cite{kingma2014auto} to compress each multi-field snapshot into a compact latent representation. The encoder $\mathcal{E}$ maps input snapshots $x \in \mathbb{R}^{C \times N_x \times N_y}$ (where $C$ is the number of physical fields and $N_x,N_y$ are the spatial grid dimensions) to a latent distribution $\mathcal{E}(x) = (\mu(x), \log \sigma^2(x))$, with latent samples obtained via the reparameterization trick: $z = \mu + \sigma \odot \epsilon$, where $\epsilon \sim \mathcal{N}(0, I)$. The decoder $\mathcal{D}$ reconstructs the input: $\hat{x} = \mathcal{D}(z)$.

The VAE must accurately reconstruct turbulence fields from their latent representations, preserving both large-scale structures and fine-scale fluctuations. Poor reconstruction would propagate errors through the temporal model and degrade prediction quality. For successful temporal modeling, the latent representations of consecutive snapshots must form smooth, learnable trajectories. Consecutive time steps should map to nearby points in latent space, avoiding discontinuous jumps that would be difficult for temporal models to predict. Distinct physical states should map to distinguishable latent representations; if all states collapse to a small region, the temporal model cannot learn meaningful dynamics, leading to a small training manifold that limits the prediction capacity.
These requirements inform our choice of loss functions for VAE training, which include not only reconstruction terms but also regularization to encourage smooth latent structure (see Section~\ref{sec:finetuning}).

\textbf{Stage 2: Temporal Dynamics in Latent Space.}
The second stage learns the temporal evolution of the compressed latent representations. Given a sequence of latent states $\{z_{t-L+1}, \ldots, z_t\}$, the temporal model predicts the next state $z_{t+1}$, which is then decoded to obtain the predicted turbulence field. We employ Long Short-Term Memory (LSTM) networks~\cite{hochreiter1997long, sherstinsky2020fundamentals} for temporal modeling, as their gating mechanism effectively captures both transient fluctuations and persistent structures (e.g., zonal flows) characteristic of plasma turbulence.

The two-stage decomposition offers practical advantages: (i) the VAE can leverage pre-trained weights from large-scale image generation models; (ii) each stage can be trained and validated independently; and (iii) the temporal model operates in a low-dimensional latent space, enabling efficient long-horizon prediction.

\subsection{Pre-trained VAE from Stable Diffusion}
\label{sec:sd_vae}

A key innovation of our approach is leveraging pre-trained VAEs rather than designing and training autoencoders from scratch. Specifically, we adopt AutoencoderKL from Stable Diffusion v1.5~\cite{rombach2022high, von-platen-etal-2022-diffusers}, though the methodology is not limited to this particular model---other pre-trained VAEs could be substituted based on the specific data characteristics or application requirements.

AutoencoderKL is the VAE component of Stable Diffusion, a latent diffusion model developed for high-resolution image synthesis. The VAE was trained to compress images into a lower-dimensional latent space while preserving visual fidelity, making it well-suited for our turbulence compression task. The AutoencoderKL architecture employs a fully convolutional encoder-decoder built upon the design principles of VQGAN~\cite{esser2021taming}. The encoder and decoder are composed of ResNet-style residual blocks~\cite{he2016deep}, where each block contains convolutional layers with skip connections---shortcut paths that add the input directly to the output, enabling effective gradient flow and allowing deeper networks to be trained stably. A key architectural component is the self-attention layer~\cite{vaswani2017attention} incorporated in the bottleneck between encoder and decoder. Self-attention computes pairwise interactions between all spatial locations, enabling the model to capture long-range spatial correlations that local convolutions cannot efficiently represent; for turbulence data, this is particularly relevant as coherent structures such as vortices and zonal flows exhibit spatial correlations spanning the entire domain. The encoder progressively downsamples the input through strided convolutions, while the decoder progressively upsamples the latent representation. By default, AutoencoderKL applies $8\times$ spatial downsampling (i.e., $N_x \times N_y \to (N_x/8) \times (N_y/8)$) with 4 latent channels; additional compression layers (e.g., linear projection) can be appended to further reduce the latent dimensionality based on application requirements. The latent space is regularized via KL divergence toward a standard normal prior, encouraging smooth, continuous latent representations---a property beneficial for temporal prediction where consecutive turbulence snapshots should map to nearby latent points.

The image-pretrained AutoencoderKL provides representations of visual patterns including edges, textures, gradients, and multi-scale structures. These low-level features transfer effectively to turbulence data, which exhibits analogous visual characteristics: coherent vortical structures, sharp gradients at interfaces, and multi-scale fluctuation patterns.

\textbf{Why pre-trained models?}
Building high-capacity VAEs from scratch faces a fundamental tension: achieving high reconstruction fidelity for complex multi-scale turbulence requires large, expressive networks with millions of parameters, yet training such networks demands correspondingly large datasets to avoid overfitting. Physics simulations, however, typically produce limited training data---our GENE dataset contains $\sim 10^3$ snapshots, orders of magnitude smaller than typical deep learning benchmarks. A small VAE trained on limited data may converge but lacks the representational capacity to capture fine-scale structures; a large VAE risks overfitting and poor generalization.

Pre-trained models resolve this tension by providing high-capacity architectures with weights already optimized on massive external datasets. The pre-trained weights encode general-purpose visual features (edges, textures, multi-scale patterns) that transfer to turbulence data, providing an initialization for fine-tuning on limited physics data. This transfer learning approach delivers the representational power of large models while requiring only modest domain-specific training data.

Beyond addressing the data limitation, pre-trained models offer practical benefits. The architecture is well-tested, eliminating manual design of network components. Pre-trained weights provide an initialization for domain-specific fine-tuning. The original 3-channel RGB architecture naturally accommodates multi-field turbulence data: for 2-field systems (e.g., HW with $n$ and $\phi$), we use the first two channels; for four-field systems such as GENE, we modify the encoder's input convolution and the decoder's output convolution from three to four channels, with one channel representing each field: $\phi$, $n$, $T_\parallel$, and $T_\perp$. Finally, the same pre-trained backbone can be fine-tuned for different turbulence systems, providing a unified methodology across plasma physics applications.

\subsection{Physics-Informed Fine-tuning}
\label{sec:finetuning}

While pre-trained weights provide an excellent starting point, fine-tuning on turbulence data is essential to achieve high reconstruction fidelity for the specific multi-scale structures present in plasma simulations. We design a composite loss function that balances reconstruction accuracy, gradient preservation, spectral fidelity, and latent regularization:
\begin{align}
\mathcal{L}_{\text{VAE}} =& \sum_{c=1}^{C} \Big[
w_{\text{recon}} \mathcal{L}_{\text{char}}(x_c, \hat{x}_c) + w_{\nabla} \norm{\nabla x_c - \nabla \hat{x}_c}^2 \Big] 
+ w_{\text{spec}} \mathcal{L}_{\text{spec}} \nonumber \\ &+ w_{\text{KL}} \, \text{KL} \left( \mathcal{N}(\mu, \operatorname{diag}(\sigma^2)) \| \mathcal{N}(0, \mathbf{I}) \right),
\label{eq:vae_loss}
\end{align}
where the sum runs over all $C$ physical field channels.

\textbf{Charbonnier reconstruction loss.}
Rather than standard MSE, we employ the Charbonnier loss $\mathcal{L}_{\text{char}}(x, \hat{x}) = M^{-1}\sum_{j=1}^{M}\sqrt{(x_j - \hat{x}_j)^2 + \epsilon^2}$ with $M$ spatial elements and $\epsilon = 10^{-3}$, which provides a smooth approximation to L1 loss that is less sensitive to outliers while preserving sharp features common in turbulence fields. This applies to all physical channels including density $n$, potential $\phi$, and temperature fields $T_\parallel$, $T_\perp$ for gyrokinetic data.

\textbf{Gradient loss.}
Turbulence fields contain sharp spatial gradients that are critical for accurate dynamics. Since spatial differentiation in Fourier space corresponds to multiplication by wavenumber $k$, the gradient loss naturally emphasizes fine-scale features (high $k$) where gradients are largest. This preserves fine-scale structures and penalizes blurring. Gradients are computed with periodic (circular) boundary conditions consistent with the simulation domain.

\textbf{Spectral loss.}
While the gradient loss emphasizes fine-scale features, it does not directly constrain the Fourier-domain energy distribution. A key limitation of previous VAE-based turbulence surrogates, including GAIT~\cite{clavier2025generative}, is their reliance on pixel-space reconstruction losses that do not explicitly enforce spectral accuracy across all scales---problematic for turbulence where accurate spectral content at both large and small scales is essential for transport statistics.
To address this, we introduce a spectral loss that directly penalizes errors in Fourier space:
\begin{equation}
\mathcal{L}_{\text{spec}} = \left\langle w(k) \left| \log\!\left(\sqrt{|\hat{X}(k)|^2+\epsilon_s}+\epsilon_s\right) - \log\!\left(\sqrt{|X(k)|^2+\epsilon_s}+\epsilon_s\right) \right| \right\rangle,
\end{equation}
where $X(k)$ and $\hat{X}(k)$ are Fourier coefficients of the original and reconstructed fields, $\epsilon_s=10^{-8}$ prevents singular logarithms, and the average is over batch samples, channels, and retained Fourier modes. The weighting function is $w(k) = (|k|/k_{\max})^\alpha$ with $\alpha = 2$. In the implementation, $k$ is formed using unit pixel spacing and $k_{\max}$ is the maximum radial Fourier frequency on the discrete grid. The logarithmic comparison penalizes relative amplitude errors, with greater weight on higher frequencies.
The exponent $\alpha$ provides a tunable physics-informed parameter that can be adjusted based on the application. For simpler systems like the HW model, we set $w_{\text{spec}} = 0$; this term becomes important for multi-scale turbulence such as the GENE data in Section~\ref{sec:gyrokinetic}.

\textbf{KL divergence.}
The KL term regularizes the latent distribution toward a standard normal prior, ensuring a smooth latent manifold suitable for temporal modeling. We report the mean KL divergence per latent dimension (i.e., averaged over the latent space) for consistent comparison across different latent dimensionalities. This regularization competes with reconstruction fidelity: stronger KL regularization encourages a more structured latent space but can degrade reconstruction quality. To balance these objectives, we set $w_{\text{KL}}$ such that the KL loss contributes at most 10\% of the total loss, allowing reconstruction terms to dominate while still maintaining sufficient latent structure. We also use a warmup schedule that gradually increases $w_{\text{KL}}$ during early training epochs to prevent posterior collapse.

\begin{figure*}[!htbp]
    \centering
    \includegraphics[width=0.9\textwidth]{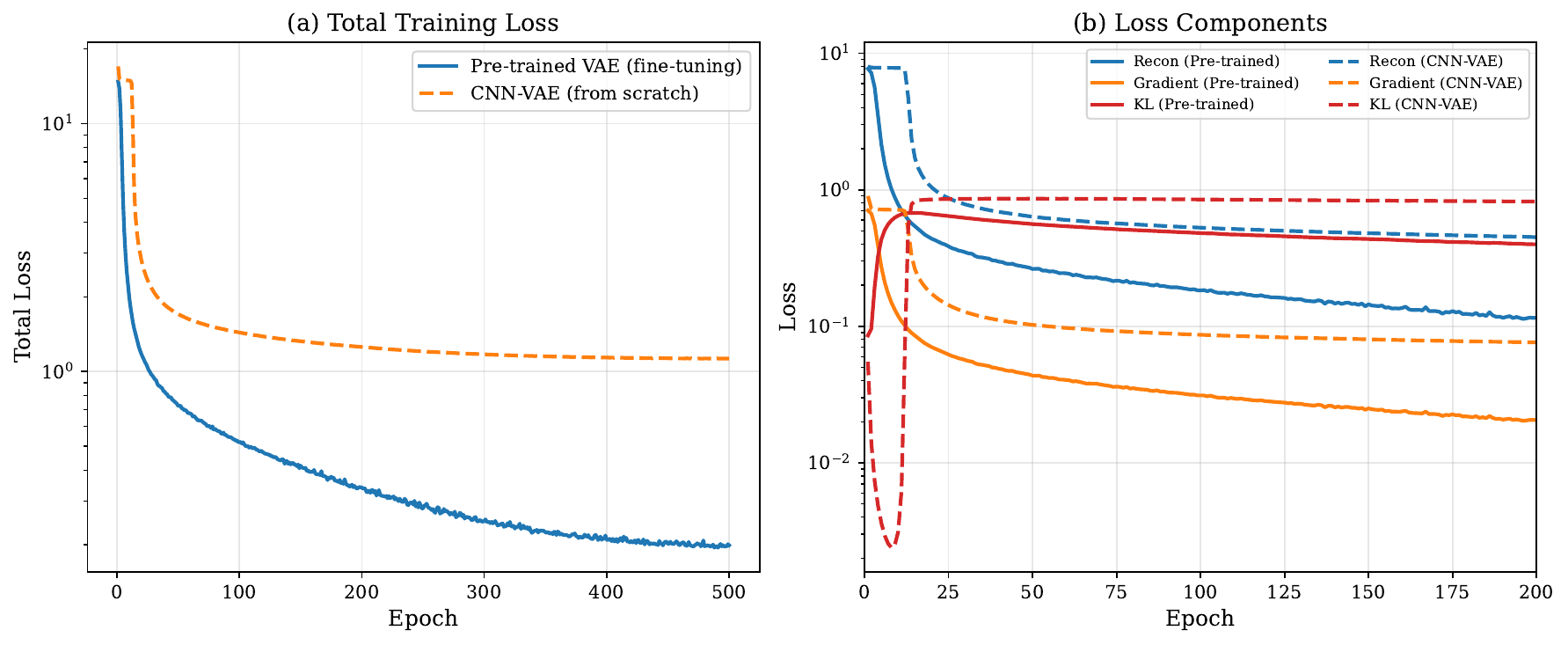}
    \caption{Training loss comparison for HW turbulence: pre-trained VAE (AutoencoderKL from Stable Diffusion v1.5, solid lines) vs.\ CNN-VAE trained from scratch (4-layer encoder/decoder, latent dim 64, dashed lines). (a) Total loss evolution. (b) Individual loss components: reconstruction (density + potential), gradient, and KL divergence. The KL loss shown is the mean per latent dimension. The pre-trained model achieves lower loss across all components, demonstrating greater representational capacity. For HW turbulence, $w_{\text{spec}} = 0$.}
    \label{fig:vae_loss}
\end{figure*}

Figure~\ref{fig:vae_loss} compares the training loss evolution for our pre-trained VAE approach versus a CNN-based VAE trained from scratch. The from-scratch CNN-VAE employs a 4-layer encoder (2$\to$32$\to$64$\to$128$\to$256 channels) with batch normalization and LeakyReLU activations, projecting to a 64-dimensional latent space, followed by a symmetric 4-layer transposed convolutional decoder. Due to its larger network size, the pre-trained VAE may require longer wall-clock time per epoch compared to the lightweight CNN-VAE. However, the key advantage lies in its representational capacity: the pre-trained model achieves a substantially lower minimum loss, indicating fast convergence in epoch steps and superior reconstruction fidelity relative to the smaller CNN-VAE over the training budget considered.

This capacity advantage is the primary benefit of the pre-trained approach. While a from-scratch model with comparable architecture size could potentially achieve similar final loss, designing and training such a model requires substantial effort in architecture selection and hyperparameter tuning. The pre-trained VAE provides this capacity immediately, and can be adapted to new datasets through fine-tuning without redesigning the core backbone.

We use a compact CNN-VAE baseline representative of the approach used in previous turbulence-surrogate work~\cite{clavier2025generative}. This provides a practical comparison for limited turbulence datasets. Since the models differ in architecture and parameter count, the comparison does not isolate the effect of pre-trained weights from model capacity.

To quantify this reconstruction advantage, we evaluate both models on 8000 HW turbulence snapshots using four complementary metrics from image quality assessment. Mean Squared Error (MSE) measures the average pixel-wise squared difference, directly quantifying reconstruction accuracy. Peak Signal-to-Noise Ratio (PSNR), defined as $10\log_{10}(\frac{x_{\max}^2}{\text{MSE}})$ in decibels, provides a logarithmic scale where each 3~dB improvement corresponds to halving the MSE; values above 30~dB typically indicate high-fidelity reconstruction. The Structural Similarity Index (SSIM) goes beyond pixel-wise comparison by evaluating luminance, contrast, and structural patterns jointly, making it sensitive to the preservation of coherent structures such as vortices and density gradients that are physically meaningful in turbulence. Finally, the Pearson correlation coefficient measures linear correspondence between original and reconstructed fields, ensuring that spatial variations are faithfully reproduced. Together, these metrics assess both point-wise accuracy and structural fidelity.

Table~\ref{tab:reconstruction} summarizes the results. The pre-trained VAE achieves 6--8$\times$ lower MSE and approximately 9~dB higher PSNR compared to the CNN-VAE, with SSIM and correlation values exceeding 0.99. These metrics confirm that the pre-trained model preserves fine-scale turbulent structures more faithfully, which is essential for accurate downstream temporal predictions.

\begin{table}[t]
\centering
\caption{Reconstruction quality metrics (mean $\pm$ std) over 8000 HW turbulence snapshots. The pre-trained VAE substantially outperforms the CNN-VAE across all metrics for both density and potential fields.}
{\footnotesize
\label{tab:reconstruction}
\begin{tabular}{llcccc}
\toprule
Model & Field & MSE $\downarrow$ & PSNR (dB) $\uparrow$ & SSIM $\uparrow$ & Correlation $\uparrow$ \\
\midrule
\multirow{2}{*}{CNN-VAE}
 & Density & 0.256 $\pm$ 0.035 & 23.7 $\pm$ 1.1 & 0.955 $\pm$ 0.007 & 0.969 $\pm$ 0.005 \\
 & Potential & 0.137 $\pm$ 0.026 & 26.2 $\pm$ 1.3 & 0.972 $\pm$ 0.006 & 0.981 $\pm$ 0.004 \\
\midrule
\multirow{2}{*}{Pre-trained}
 & Density & \textbf{0.041 $\pm$ 0.065} & \textbf{32.5 $\pm$ 2.7} & \textbf{0.993 $\pm$ 0.009} & \textbf{0.995 $\pm$ 0.008} \\
 & Potential & \textbf{0.019 $\pm$ 0.050} & \textbf{35.7 $\pm$ 2.7} & \textbf{0.996 $\pm$ 0.007} & \textbf{0.997 $\pm$ 0.007} \\
\bottomrule
\end{tabular}}
\end{table}

\subsection{Temporal Dynamics in Latent Space}

The temporal model learns to predict the evolution of latent representations. Given a sequence of encoded latent states $\{z_{t-L+1}, \ldots, z_t\}$, the model predicts the next latent state $\hat{z}_{t+1}$, which is then decoded to obtain the predicted turbulence field. LSTM networks~\cite{hochreiter1997long} are recurrent neural networks designed to capture long-range temporal dependencies. Standard recurrent networks suffer from vanishing gradients when learning dependencies over many time steps; LSTMs address this through a gating mechanism that selectively retains or discards information. Each LSTM cell maintains a cell state $c_t$ that acts as a memory, regulated by three gates: the forget gate determines what information to discard from the previous cell state, the input gate controls what new information to store, and the output gate determines what to expose as the hidden state $h_t$. This gating mechanism enables LSTMs to learn both fast fluctuations and slowly-evolving structures characteristic of plasma turbulence, such as zonal flows that persist over many turbulence correlation times.

\textbf{Convolutional LSTM for spatial latents.}
For the GENE application, the VAE produces latent representations that retain spatial structure ($z \in \mathbb{R}^{4 \times 16 \times 6}$). Applying a standard fully-connected LSTM would require flattening this tensor to a vector, destroying the spatial correlations between neighboring latent pixels. Convolutional LSTM (ConvLSTM)~\cite{shi2015convolutional} addresses this limitation by replacing the matrix multiplications in standard LSTM with convolution operations:
\begin{equation}
    h_t, c_t = \text{ConvLSTM}(x_t, h_{t-1}, c_{t-1}; W_*),
\end{equation}
where $h_t$ and $c_t$ are now 3D tensors preserving spatial dimensions, and $W_*$ denotes convolutional filter weights for the gates. The key insight is that convolutions are local operations: each spatial location in the hidden state is updated based on its neighborhood in both the input and previous hidden state. This locality is well-suited for turbulence, where dynamics are governed by local interactions---eddies influence their neighbors, and structures propagate spatially over time. ConvLSTM can thus learn spatially-varying temporal patterns, such as how turbulent fluctuations at one location affect neighboring regions.

Our ConvLSTM implementation uses 3 stacked layers with 256 hidden channels and $3 \times 3$ convolutional kernels, followed by a $1 \times 1$ convolution that projects back to the latent channel dimension. For simpler cases with lower-dimensional latent vectors, standard fully-connected LSTM layers can be used instead.
Training minimizes a combined loss with smoothness regularization:
\begin{equation}
\mathcal{L}_{\text{LSTM}} = \text{MSE}(\hat{z}_{t+1}, z_{t+1}) + \lambda_{\text{smooth}} \sum_{t} \norm{\hat{z}_{t+1} - \hat{z}_t}^2,
\end{equation}
where the smoothness term ($\lambda_{\text{smooth}} = 0.01$) encourages temporally coherent predictions. Latent sequences are normalized per-channel before training. We use AdamW optimizer with learning rate $10^{-4}$, gradient clipping, and early stopping based on validation loss. The sequence length $L = 50$ provides sufficient temporal context for capturing turbulence dynamics.

\textbf{Manifold consistency error.}
A fundamental challenge in latent space modeling is \emph{distribution shift}: during autoregressive rollout, prediction errors accumulate and can push latent states off the valid manifold learned by the VAE. Unlike single-step prediction where errors remain bounded, autoregressive inference feeds each prediction back as input, allowing small deviations to compound over hundreds or thousands of steps. If predicted latents drift into regions where the VAE decoder was never trained, the reconstructed fields can exhibit artifacts or unphysical behavior.

To monitor this, we define a \emph{manifold consistency error} that measures encode--decode consistency of predicted latents:
\begin{equation}\label{eq:E_main}
    E_{\text{mani}} = \frac{1}{d_z}\| \hat{z} - \mathcal{E}_\mu(\mathcal{D}(\hat{z})) \|^2,
\end{equation}
where $d_z$ is the number of latent elements and $\mathcal{E}_\mu$ denotes deterministic encoding using the posterior mean, with the same VAE scaling as $\hat{z}$. The metric is evaluated after undoing the per-channel normalization used by the temporal model. It measures the mean squared discrepancy between a predicted latent and its decoded and re-encoded representation. Re-encoding is used only for this diagnostic; the predicted latent itself is fed back during rollout. A stable value indicates stable encode--decode consistency, but does not by itself establish physical accuracy or membership in the training distribution.

\section{Validation on HW Turbulence}
\label{sec:hw_validation}

We validate the surrogate model through statistical diagnostics that characterize turbulence across spatial scales, temporal frequencies, and multi-field correlations. These metrics, applied consistently throughout this work for both HW and GENE turbulence datasets, provide complementary perspectives on whether the surrogate captures the essential physics.

The spatial amplitude spectrum $S_f(k) = \langle | \mathcal{F}_{xy}[\delta f](\mathbf{k}) | \rangle_{|\mathbf{k}|=k, t}$ characterizes fluctuation amplitudes across spatial scales, where $k = \sqrt{k_x^2 + k_y^2}$ is the radial wavenumber. The temporal amplitude spectrum $\Omega_f(\omega) = \langle | \mathcal{F}_t[\delta f](\omega) | \rangle_{\mathbf{x}}$ characterizes the frequency content of turbulent dynamics.
Proper Orthogonal Decomposition (POD)~\cite{beyer2000proper,futatani2009spatiotemporal} provides an energy-ranked basis for turbulent fields via singular value decomposition of the snapshot matrix: $\mathbf{A} = \mathbf{U} \boldsymbol{\Sigma} \mathbf{V}^T$, with POD spectrum $\sigma_\ell = \Sigma_{\ell\ell}$ where $\sigma_\ell^2$ is proportional to the energy in the $\ell$-th mode.
{Temporal autocorrelation} $$C_f(\tau) = \langle \delta f(\mathbf{x},t) \, \delta f(\mathbf{x},t+\tau) \rangle / \langle \delta f^2 \rangle$$ quantifies how rapidly fluctuations decorrelate in time, and the cross-correlation $$C_{AB}(\tau) = \langle \delta A(\mathbf{x},t) \, \delta B(\mathbf{x},t+\tau) \rangle / (\sigma_A \sigma_B)$$ captures phase relationships between fields that govern transport.

\begin{figure*}[!htbp]
    \centering
    \includegraphics[width=\textwidth]{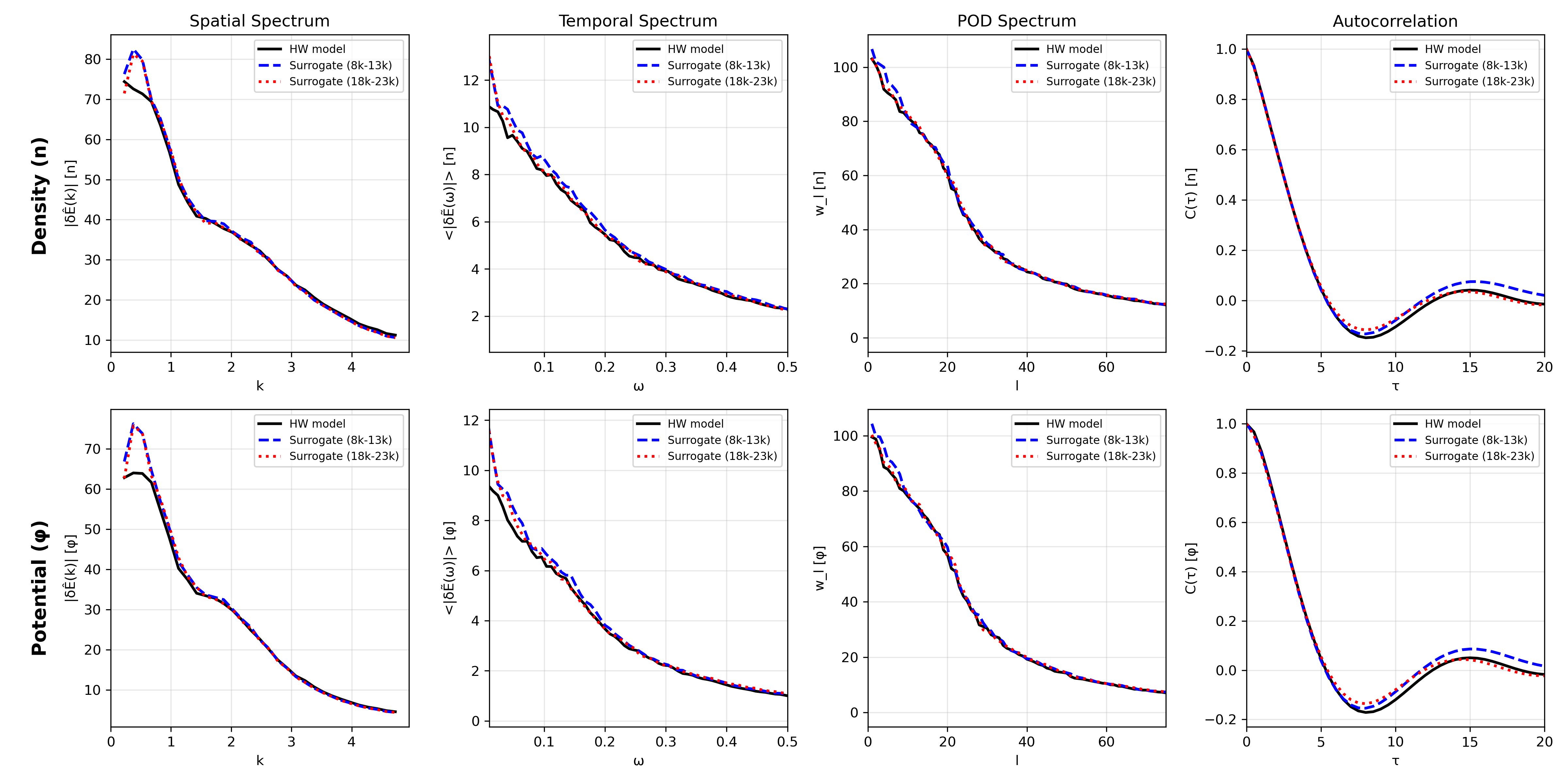}
    \caption{Statistical validation of PreVAE-Turb on HW turbulence for density $n$ (top row) and potential $\phi$ (bottom row). From left to right: spatial spectrum $S_f(k)$, temporal spectrum $\Omega_f(\omega)$, POD spectrum $\sigma_\ell$, and temporal autocorrelation $C_f(\tau)$. Black solid: HW ground truth. Blue dashed: PreVAE-Turb prediction $t\in [8000,13000]$. Red dotted: PreVAE-Turb prediction $t\in [18000, 23000]$. PreVAE-Turb accurately reproduces the statistical properties across both fields and different prediction windows.}
    \label{fig:hw_statistics}
\end{figure*}

\begin{figure*}[!htbp]
    \centering
    \includegraphics[width=\textwidth]{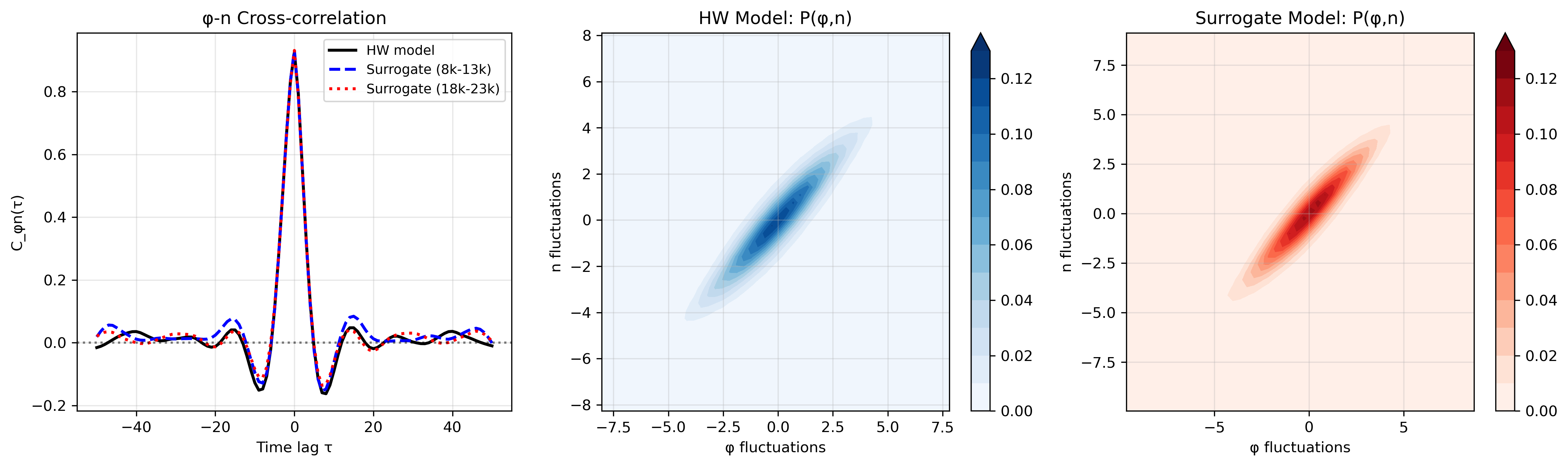}
    \caption{Cross-field correlations for HW turbulence. Left: $\phi$-$n$ cross-correlation $C_{\phi n}(\tau)$ showing agreement between HW model (black) and PreVAE-Turb predictions at different time windows (blue dashed: $t\in[8000, 13000]$, red dotted: $t\in [18000, 23000])$. Middle: joint probability distribution $P(\phi, n)$ for HW ground truth. Right: joint probability distribution $P(\phi, n)$ for PreVAE-Turb prediction. PreVAE-Turb preserves the characteristic tilted elliptical structure indicating the phase relationship between potential and density fluctuations.}
    \label{fig:hw_crossfield}
\end{figure*}

Figure~\ref{fig:hw_statistics} validates PreVAE-Turb on HW turbulence through these metrics. Throughout this paper, time is normalized by the temporal resolution $\Delta t$ of each dataset, so $t$ denotes the snapshot index rather than physical time; that is, $t = k$ refers to the $k$-th snapshot, corresponding to physical time $k \Delta t$. PreVAE-Turb is initialized with ground-truth data and then run autoregressively, with each prediction window spanning 5000 rollout steps. The first window ($t \in [8000, 13000]$) overlaps with the training period, while the second window ($t \in [18000, 23000]$) extends well beyond the training data, testing long-horizon generalization. Spatial and temporal amplitude spectra show overall agreement across scales, POD singular value decay matches the reference, and temporal autocorrelation captures the characteristic decorrelation time. Beyond single-field statistics, Figure~\ref{fig:hw_crossfield} examines cross-field correlations that govern transport. The cross-correlation $C_{n\phi}(\tau)$ captures the phase relationship between density and potential that determines particle flux, while the joint probability distribution $P(n, \phi)$ reveals the statistical coupling structure. PreVAE-Turb accurately reproduces the partial correlation characteristic of the HW system, confirming that the model captures the density-potential coupling mechanisms essential for transport predictions.

\section{Application to Gyrokinetic Turbulence}
\label{sec:gyrokinetic}

Gyrokinetic simulations represent the state-of-the-art for first-principles modeling of plasma turbulence in magnetic confinement fusion devices. The Gyrokinetic Electromagnetic Numerical Experiment (GENE)~\cite{jenko2000electron,gorler2011gene} solves the nonlinear gyrokinetic Vlasov-Maxwell system in five-dimensional phase space $(x, y, z, v_\parallel, \mu)$, capturing the full kinetic dynamics of turbulent fluctuations across multiple scales. Gyrokinetic theory provides a rigorous reduction from the 6D Vlasov equation by averaging over the fast gyromotion and the resulting 5D problem is tractable.  However, gyrokinetic simulations remain computationally prohibitive for many applications of interest.

The computational cost of gyrokinetic simulations motivates the development of efficient surrogate models. A trained surrogate that captures the essential turbulence dynamics could enable rapid parameter scans, uncertainty quantification, and real-time applications that are currently intractable with direct numerical simulation. The present work represents an initial step towards such capabilities.  Having validated our framework on HW turbulence, we now demonstrate its extension to multi-field gyrokinetic turbulence from GENE simulations.

\subsection{GENE Dataset and Multi-field Extension}

Gyrokinetic simulations evolve a 5D distribution function in time.  In practice, many quantities of interest can be extracted from a reduction of this dataset.  Here, we analyze four fields: the electrostatic potential $\phi$, and three moments of the ion distribution function---the density $n_i$ and the parallel/perpendicular temperatures $T_{\parallel,i}$, $T_{\perp,i}$.  Moreover, we further reduce the data to 2 dimensions by considering only the data at the outboard midplane---i.e., a single point in the parallel (to the magnetic field) dimension.  The resulting dataset comprises snapshots of shape $(4 \times 128 \times 48)$.

Prior to training, each field is normalized independently to zero mean and unit variance across the full dataset. This channel-wise normalization ensures balanced learning across fields with different physical units and magnitudes, and is standard practice for multi-channel neural network training. All validation metrics (spectra, cross-correlations, fluxes) are computed on the normalized data; since both GENE ground truth and surrogate predictions use the same normalization, relative comparisons and correlation-based metrics remain valid.

The training data is generated by a nonlinear GENE simulation defined by the following parameters: normalized inverse gradient scale lengths $a/L_{Ti} = a/L_{Te} = 6.96$, $a/L_{n_i} = a/L_{ne} = 2.22$.  We use simple $\hat{s}-\alpha$ geometry defined by magnetic shear $\hat{s} = 0.796$, safety factor $q = 1.4$, inverse aspect ratio $\epsilon = 0.18$, and $\alpha = 0$.  This is similar to the canonical cyclone base case, except that kinetic electrons are included.  The grid resolution is defined by 128 Fourier modes in the radial direction (ranging from $-k_{x,\max}$ to $k_{x,\max}$), 24 Fourier modes in the binormal direction (ranging from zero to $k_{y,max}$), 16 parallel grid points, and $(32,8)$ grid points in the parallel and perpendicular velocity coordinates $(v_{||},\mu)$, respectively.  The box sizes are $126 \rho_s$ in both the radial and binormal directions.  

We expand the pre-trained VAE from 3-channel RGB to 4-channel by modifying encoder/decoder convolution layers: The first three channels reuse pre-trained weights directly, and the fourth channel is initialized by averaging RGB weights.
The encoder maps $\mathbf{x} \in \mathbb{R}^{4 \times 128 \times 48} \to \mathbf{z} \in \mathbb{R}^{4 \times 16 \times 6}$, achieving a compression ratio of 64. The higher-dimensional latent space (compared to HW) retains spatial structure, motivating the use of ConvLSTM for temporal dynamics as described in Section~\ref{sec:method}. The loss function follows Eq.~\eqref{eq:vae_loss} with all components (Charbonnier reconstruction, gradient, spectral, and KL terms), and the manifold consistency error Eq.~\eqref{eq:E_main} is monitored during autoregressive rollout to assess encode--decode consistency.

\subsection{Comparison with CNN-VAE Trained from Scratch}

As with the HW case (Section~\ref{sec:finetuning}), we compare against a CNN-VAE baseline trained from scratch to compare the pre-trained backbone with a compact from-scratch alternative. The GENE CNN-VAE employs a convolutional encoder with three downsampling stages ($4 \to 64 \to 128 \to 256 \to 256$ channels) using residual blocks with GroupNorm and SiLU activations, producing a latent representation of shape $(4 \times 16 \times 6)$---the same spatial latent dimensions as the pre-trained VAE---followed by a symmetric decoder with transposed convolutions. Both models are trained for 1000 epochs using identical loss functions (Eq.~\eqref{eq:vae_loss}) with the same weights: $w_{\text{recon}} = 1.0$, $w_\nabla = 1.0$, $w_{\text{spec}} = 0.05$, $w_{\text{KL}} = 3 \times 10^{-5}$. Training uses AdamW optimizer with learning rate $10^{-4}$, batch size 16, and cosine annealing schedule. The rationale for comparing against a smaller CNN-VAE rather than scaling it to match the pre-trained model's parameter count was discussed in Section~\ref{sec:finetuning}.

\begin{figure*}[!htbp]
    \centering
    \includegraphics[width=0.9\textwidth]{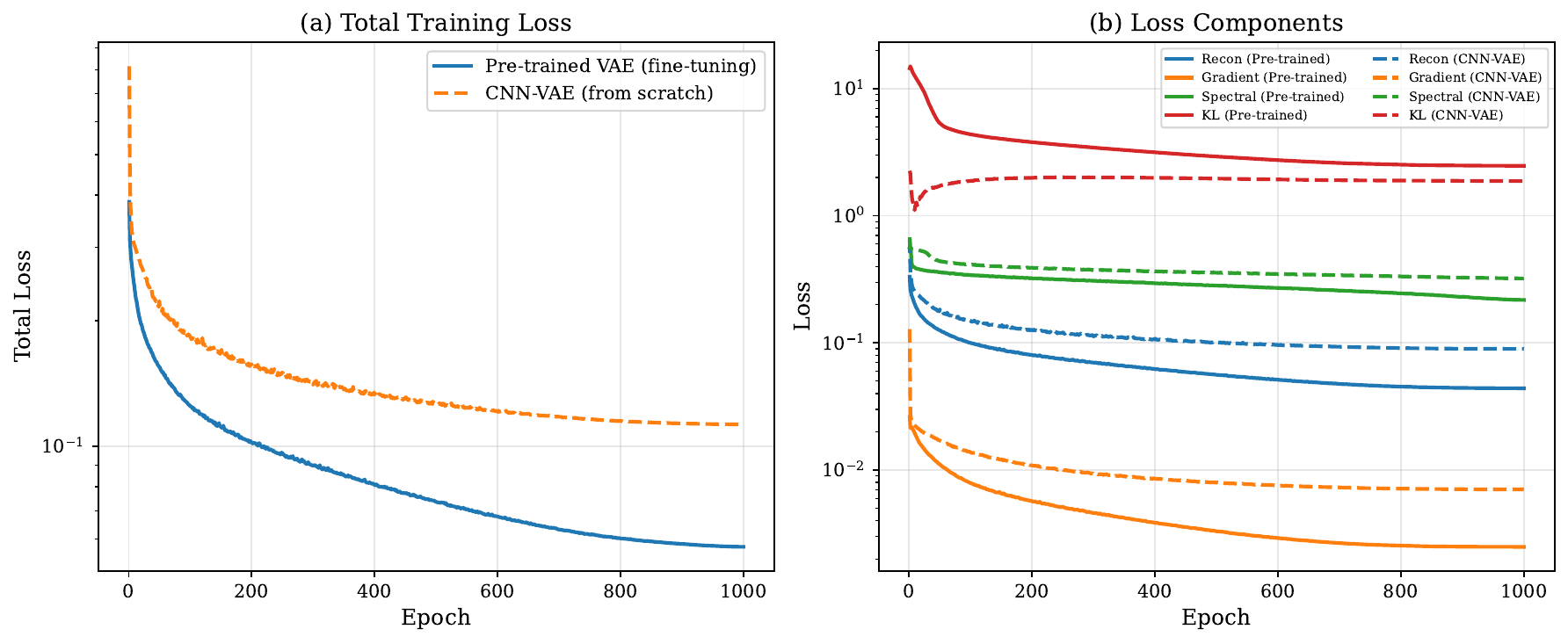}
    \caption{VAE training loss comparison for GENE turbulence: pre-trained VAE (solid lines) vs.\ CNN-VAE trained from scratch (dashed lines). Left: total loss evolution. Right: individual loss components (reconstruction, gradient, spectral, and KL divergence). The pre-trained model achieves lower loss across all reconstruction-related terms while maintaining comparable KL regularization.}
    \label{fig:gene_vae_loss}
\end{figure*}

Figure~\ref{fig:gene_vae_loss} compares the training loss evolution for both approaches. The pre-trained VAE achieves substantially lower total loss and faster convergence in terms of epochs. The reconstruction, gradient, and spectral loss components all show significant improvement for the pre-trained model, while the KL divergence terms remain comparable---this is expected since the KL term regularizes the latent space structure rather than reconstruction quality. These results mirror the HW comparison (Figure~\ref{fig:vae_loss}) and confirm that the pre-trained approach provides superior representational capacity for multi-field gyrokinetic data.

\subsection{Validation on GENE Turbulence}

We apply the validation metrics defined in Section~\ref{sec:hw_validation} to assess the GENE surrogate. The dataset comprises $N_t = 1400$ snapshots at temporal resolution $\Delta t = 0.431$ (in normalized GENE units), on a $128 \times 48$ spatial grid. The VAE and ConvLSTM are trained on the first 1200 snapshots, with the remainder reserved for validation. Following the time normalization convention in Section~\ref{sec:hw_validation}, $t$ denotes the snapshot index; for example, $t = 1200$ refers to the 1200th snapshot, corresponding to physical time $1200 \times \Delta t \approx 517$ in normalized GENE units.

An important distinction must be made between training and inference. During inference, PreVAE-Turb operates autoregressively: it predicts the next latent state and feeds that prediction directly back to the ConvLSTM, without access to ground truth data. The resulting latent trajectory is decoded to obtain the physical fields. For GENE, the encoder uses the deterministic posterior mean, multiplied by the VAE scaling factor. To initialize the LSTM hidden states, we use a warmup phase where the first 400 ground-truth snapshots are encoded and fed through the LSTM sequentially. After this warmup ($t \geq 400$), PreVAE-Turb runs fully autonomously, using only its own predictions as input.

\begin{figure*}[!htbp]
    \centering
    \includegraphics[width=\textwidth]{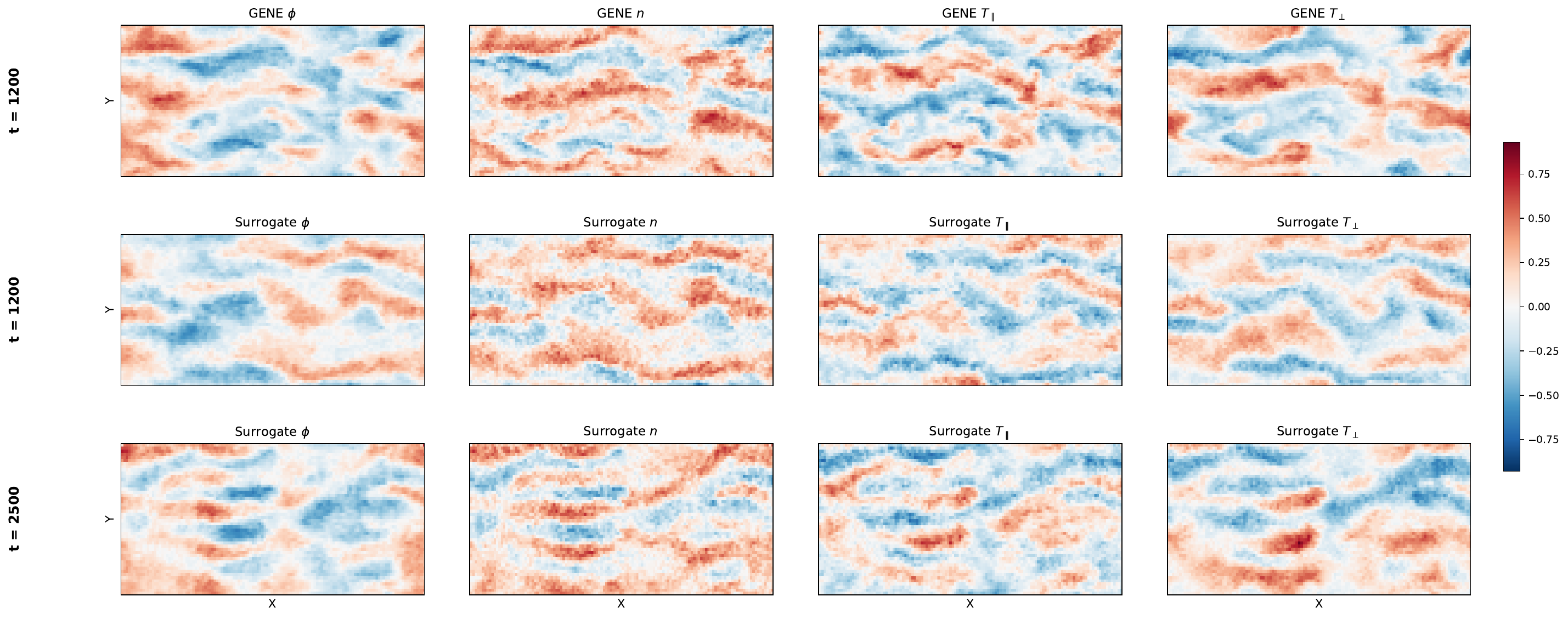}
    \caption{Snapshot comparison for GENE turbulence fields. Top row: GENE ground truth at $t = 1200$. Middle row: PreVAE-Turb prediction at $t = 1200$. Bottom row: PreVAE-Turb prediction at $t = 2500$ (long-horizon extrapolation). Columns from left to right: electrostatic potential $\phi$, density $n$, parallel temperature $T_\parallel$, and perpendicular temperature $T_\perp$. PreVAE-Turb accurately reproduces coherent structures at $t = 1200$ and maintains stable turbulence patterns at $t = 2500$, demonstrating long-term prediction stability.}
    \label{fig:gene_snapshot}
\end{figure*}

We compare GENE ground truth against PreVAE-Turb predictions over two time windows: $t \in [400, 900]$ and $t \in [2400, 2900]$. Although the first window overlaps with the training data temporally, the PreVAE-Turb predictions in this window are \emph{not} reconstructions of training data---they are autoregressive rollouts that accumulate prediction errors over 500 steps. The second window extends far beyond the training horizon, testing long-term statistical stability. Since reference data are unavailable at those later times, its statistics are compared with the earlier GENE reference window.
Figure~\ref{fig:gene_snapshot} compares representative snapshots at $t = 1200$ (within the prediction region with available ground truth) and $t = 2500$ (long-horizon extrapolation beyond training data), demonstrating that PreVAE-Turb preserves the spatial structure and amplitude of turbulent fluctuations across all four fields. Even at $t = 2500$, far beyond the training window, PreVAE-Turb maintains stable and physically plausible turbulence patterns.

\begin{figure*}[!htbp]
    \centering
    \includegraphics[width=\textwidth]{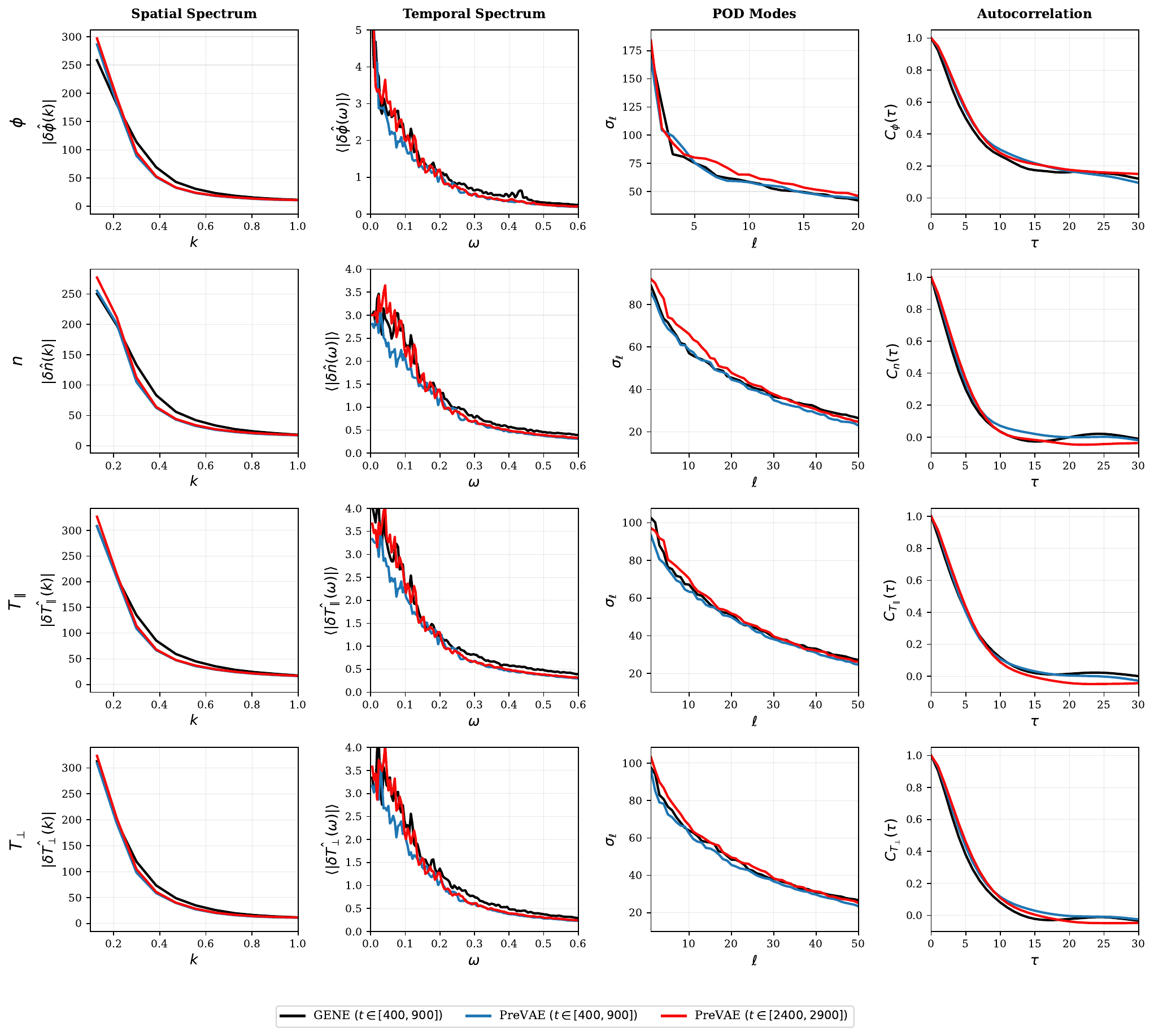}
    \caption{Validation metrics for PreVAE-Turb across all four fields ($\phi$, $n$, $T_\parallel$, $T_\perp$). Columns: spatial spectrum $S_f(k)$, temporal spectrum $\Omega_f(\omega)$, POD spectrum $\sigma_\ell$, and autocorrelation $C_f(\tau)$. Black: GENE ground truth, $t\in[400,900]$. Blue: PreVAE-Turb, $t\in[400,900]$. Red: PreVAE-Turb, $t\in[2400,2900]$.}
    \label{fig:gene_validation}
\end{figure*}

Figure~\ref{fig:gene_validation} presents the validation results across all four fields. The spatial spectra (first column) reproduce the overall scale dependence, with an overshoot at the lowest wavenumbers and an underestimate at intermediate wavenumbers. The temporal spectra (second column) demonstrate that PreVAE-Turb captures both low-frequency coherent structures and higher-frequency fluctuations, with similar spectral shapes in both rollout windows. The POD spectra (third column) match in both slope and magnitude, indicating that PreVAE-Turb preserves the hierarchy of coherent structures. The autocorrelation functions (fourth column) confirm that the temporal model learns the correct decorrelation timescales.

\begin{figure*}[!htbp]
    \centering
    \includegraphics[width=\textwidth]{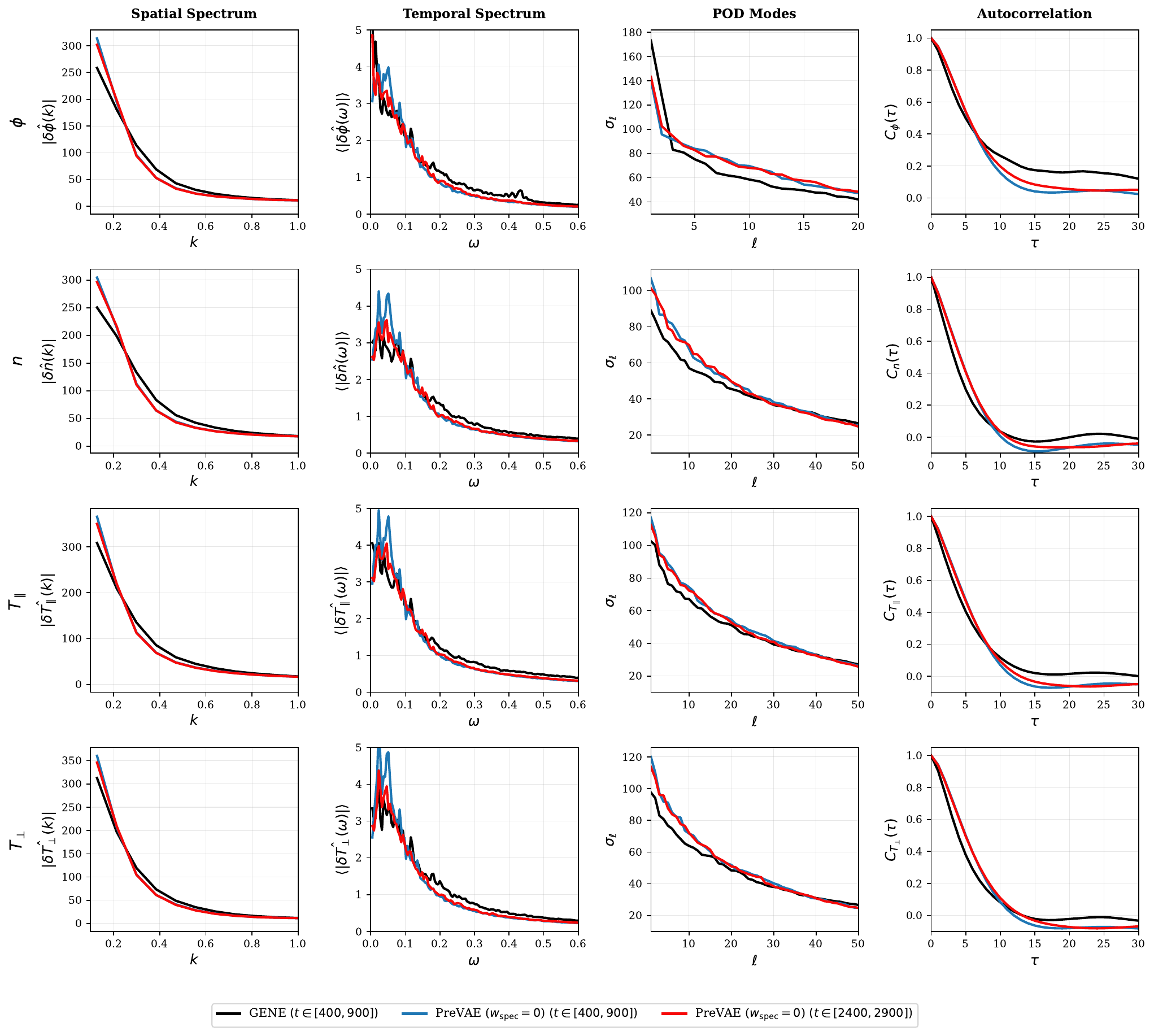}
    \caption{Ablation study: validation metrics for PreVAE-Turb trained without spectral loss ($w_{\text{spec}} = 0$). Compared to Figure~\ref{fig:gene_validation}, differences are visible in POD spectra and autocorrelation, while the spatial spectra remain broadly similar.}
    \label{fig:spectral_ablation}
\end{figure*}

To demonstrate the importance of the spectral loss (Eq.~\ref{eq:vae_loss}), we conduct an ablation study by training PreVAE-Turb with $w_{\text{spec}} = 0$ while keeping all other hyperparameters identical. Figure~\ref{fig:spectral_ablation} compares the validation metrics between the full PreVAE-Turb model and the ablated version without spectral loss.
The comparison shows differences in POD spectra and autocorrelation structure, while the spatial spectra remain broadly similar. These results suggest that the spectral term affects downstream rollout statistics in the runs considered; they do not establish that it is necessary for accurate spectra in all cases.

\begin{figure}[!htbp]
    \centering
    \includegraphics[width=0.5\columnwidth]{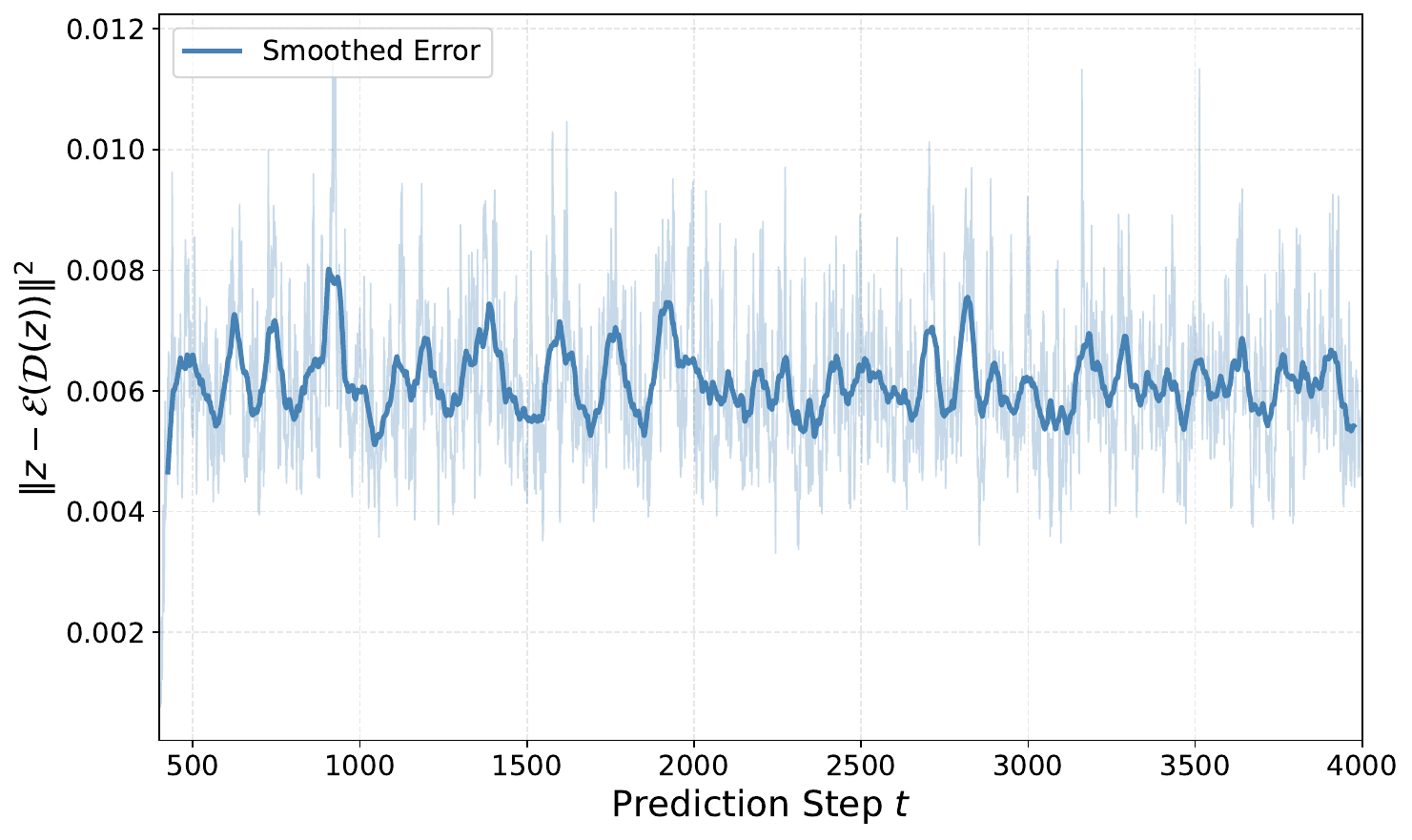}
    \caption{Manifold consistency error $E_{\text{mani}}$ (Eq.~\ref{eq:E_main}) during autoregressive rollout, averaged over latent elements. The error remains around $0.006$, indicating stable encode--decode consistency over the plotted interval.}
    \label{fig:manifold_consistency}
\end{figure}

To monitor latent consistency, we measure the manifold consistency error (Eq.~\ref{eq:E_main}) for each predicted latent during autoregressive rollout. Figure~\ref{fig:manifold_consistency} shows this error over the long rollout. The error remains around $0.006$, with no sustained upward trend, indicating that encode--decode consistency does not progressively deteriorate. Physical fidelity is assessed separately through the field and transport diagnostics.

\begin{figure*}[!htbp]
    \centering
    \includegraphics[width=\textwidth]{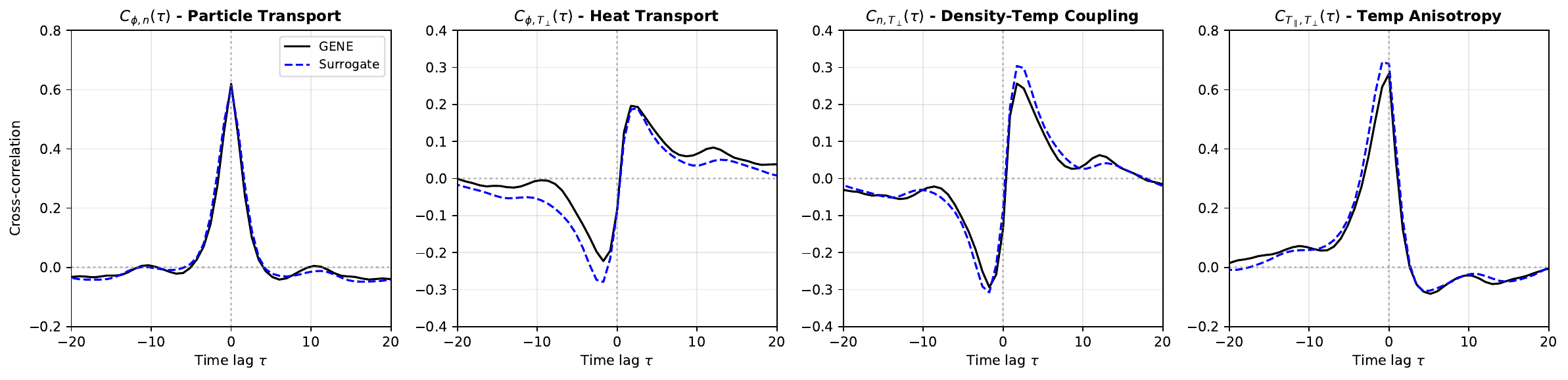}
    \caption{Cross-correlation functions between GENE field pairs. From left to right: $\phi$-$n$ (particle transport), $\phi$-$T_\perp$ (perpendicular heat transport), $n$-$T_\perp$ (density-temperature coupling), and $T_\parallel$-$T_\perp$ (temperature anisotropy). Black lines: GENE ground truth. Blue lines: PreVAE-Turb prediction. The agreement in cross-correlations demonstrates that PreVAE-Turb preserves the multi-field coupling structure essential for transport predictions.}
    \label{fig:gene_crosscorr}
\end{figure*}

Beyond single-field statistics, cross-correlations between fields govern turbulent transport. For GENE's four-field system, we evaluate $C_{\phi n}$ (particle transport), $C_{\phi T_\perp}$ (perpendicular heat flux), $C_{n T_\perp}$ (density-temperature coupling), and $C_{T_\parallel T_\perp}$ (temperature anisotropy). Figure~\ref{fig:gene_crosscorr} shows that PreVAE-Turb accurately reproduces the correlation structure, including peak values and characteristic lag times. The $\phi$-$n$ correlation shows strong positive coupling at zero lag, consistent with electrostatic ITG-driven turbulence.

\begin{figure*}[!htbp]
    \centering
    \includegraphics[width=\textwidth]{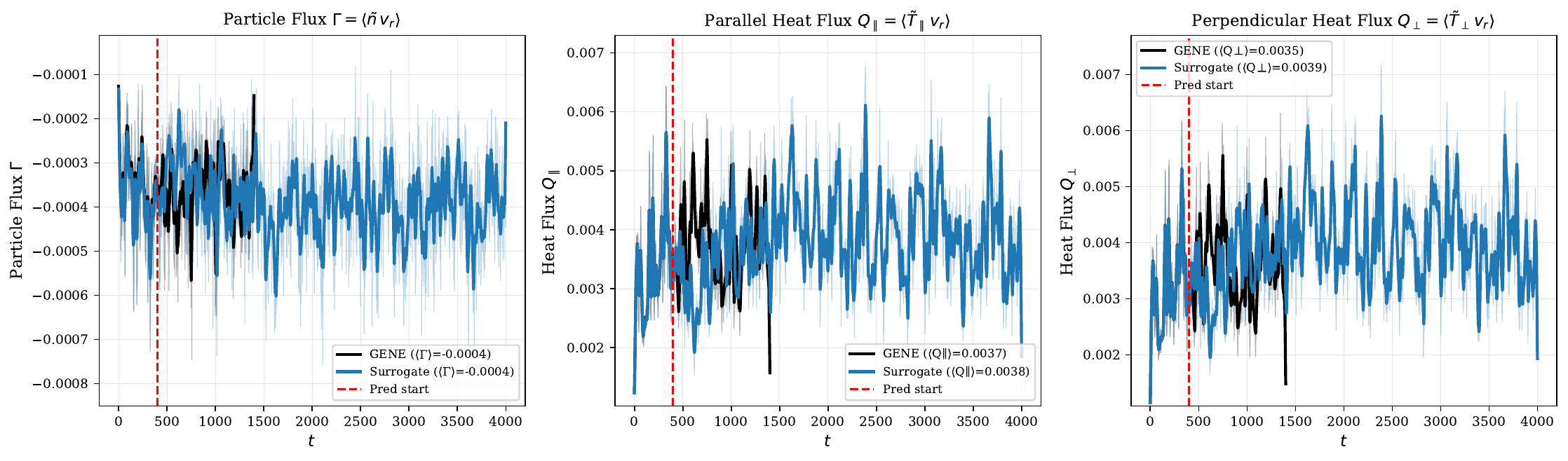}
    \caption{Temporal evolution of normalized midplane transport proxies. Left: particle flux $\Gamma = \langle \tilde{n} \, v_r \rangle$. Middle: parallel heat flux $Q_\parallel = \langle \tilde{T}_\parallel \, v_r \rangle$. Right: perpendicular heat flux $Q_\perp = \langle \tilde{T}_\perp \, v_r \rangle$. Black: GENE ground truth ($t = 0$--$1400$). Blue: PreVAE-Turb prediction ($t = 0$--$4000$). Faint lines show raw data; bold lines show smoothed (moving average) values. The red dashed line at $t = 400$ marks the start of autoregressive prediction; before this, the outputs are VAE reconstructions of ground truth latents (warmup period).}
    \label{fig:gene_flux}
\end{figure*}

The ultimate validation for a turbulence surrogate is its ability to predict transport fluxes, which determine confinement performance in fusion devices. We compute normalized midplane transport proxies: the particle-flux proxy $\Gamma = \langle \tilde{n} \, v_r \rangle$ and temperature-flux proxies $Q_\parallel = \langle \tilde{T}_\parallel \, v_r \rangle$, $Q_\perp = \langle \tilde{T}_\perp \, v_r \rangle$, where $v_r = -\partial \phi / \partial y$ is the radial $E \times B$ velocity proxy in the adopted sign convention and tildes denote fluctuations about the spatial mean. Because these quantities use standardized fields at a single outboard-midplane slice, they are not full flux-surface-averaged GENE particle and heat fluxes. We retain $\Gamma$, $Q_\parallel$, and $Q_\perp$ as shorthand for these diagnostics below. Figure~\ref{fig:gene_flux} shows the temporal evolution of these transport quantities. During the warmup period ($t < 400$), the outputs are VAE reconstructions of ground truth latents; after the warmup ($t \geq 400$, marked by the red dashed line), PreVAE-Turb runs fully autoregressively. PreVAE-Turb (blue) tracks the GENE ground truth (black, available for $t < 1400$) in the overlapping region, capturing both the mean flux levels and the characteristic fluctuation amplitudes. Beyond the ground truth window, PreVAE-Turb maintains statistically consistent flux behavior up to $t = 4000$, demonstrating long-horizon stability. The fluxes have comparable mean levels and fluctuation amplitudes, although differences in their distributions remain, as examined below.

To provide a more quantitative assessment of flux statistics, we analyze the probability density function (PDF) and temporal autocorrelation of the flux time series. Since the flux at each time step is a spatially-averaged scalar, $\Gamma(t) = \langle \tilde{n} \, v_r \rangle_{x,y}$, we obtain a one-dimensional time series from which we can compute statistical distributions. The PDF characterizes the distribution of flux values, while the temporal autocorrelation
\begin{equation}
    C_\Gamma(\tau) = \frac{\langle \delta\Gamma(t) \, \delta\Gamma(t+\tau) \rangle_t}{\langle (\delta\Gamma)^2 \rangle_t}, \qquad \delta\Gamma(t) = \Gamma(t)-\langle\Gamma\rangle_t,
\end{equation}

quantifies the correlation timescale of flux fluctuations---physically, this represents the characteristic duration of transport bursts in the turbulence.

\begin{figure*}[!htbp]
    \centering
    \includegraphics[width=\textwidth]{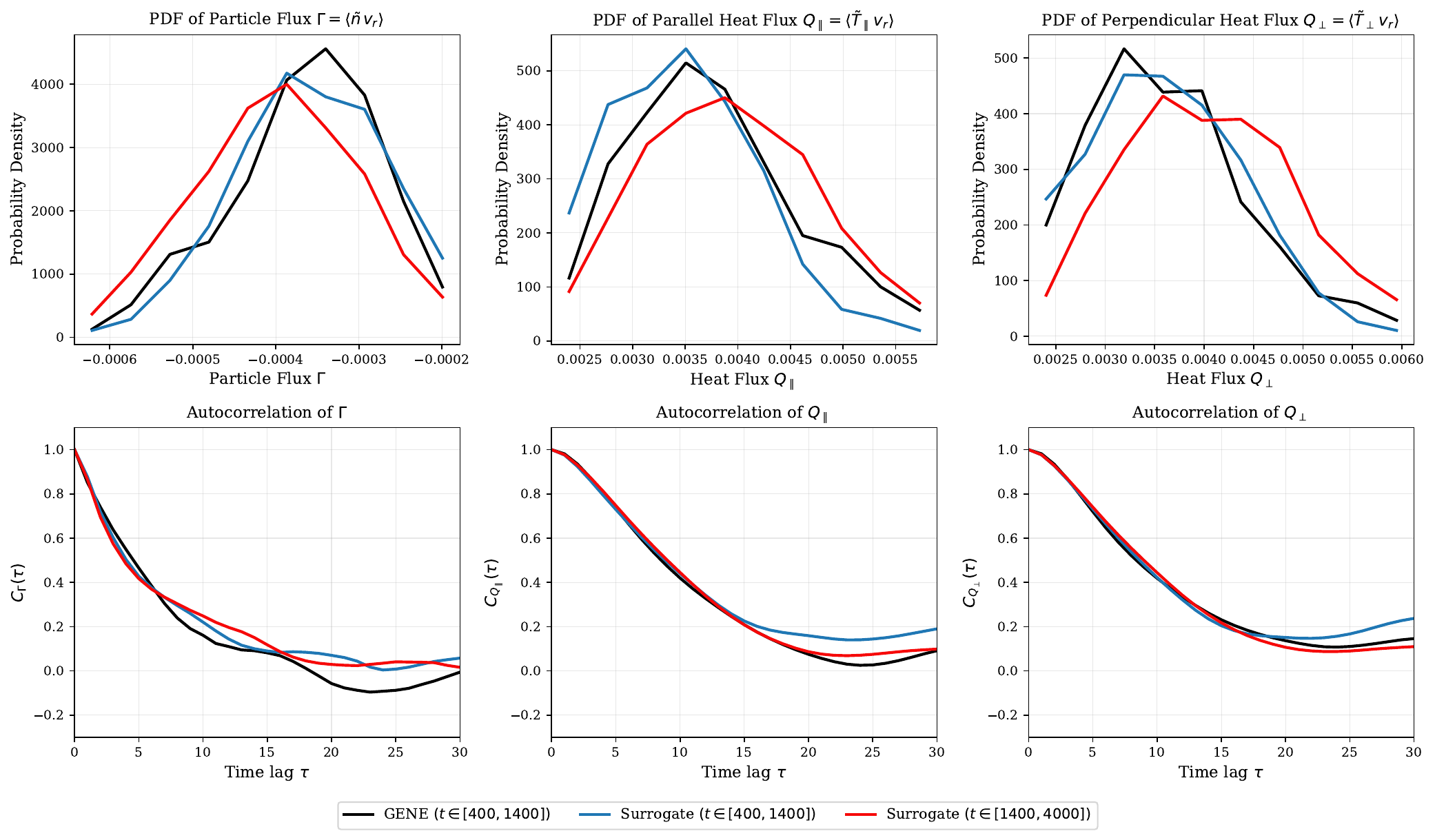}
    \caption{Normalized midplane transport-proxy probability density functions (top row) and temporal autocorrelations (bottom row) for particle flux $\Gamma$ (left), parallel heat flux $Q_\parallel$ (middle), and perpendicular heat flux $Q_\perp$ (right). Black: GENE ground truth ($t \in [400, 1400]$). Blue: PreVAE-Turb prediction ($t \in [400, 1400]$). Red: PreVAE-Turb long-horizon prediction ($t \in [1400, 4000]$). PreVAE-Turb reproduces the flux distributions and decorrelation timescales, with visible shifts in the long-horizon PDF peaks.}
    \label{fig:gene_flux_pdf}
\end{figure*}

Figure~\ref{fig:gene_flux_pdf} compares these statistics across three cases: GENE ground truth ($t \in [400, 1400]$), PreVAE-Turb in the prediction window with available ground truth ($t \in [400, 1400]$), and PreVAE-Turb in a longer-horizon prediction window without ground truth ($t \in [1400, 4000]$). The flux PDFs (top row) show that PreVAE-Turb accurately reproduces the shape and width of the GENE distributions in the early prediction window where ground truth is available. For the longer-horizon prediction ($t \in [1400, 4000]$), shifts in the PDF peaks and changes in the distribution widths are observed, particularly for the temperature-flux diagnostics. The autocorrelation functions (bottom row) demonstrate that PreVAE-Turb captures the correct decorrelation timescale across all three fluxes, with similar initial decay times for GENE and PreVAE-Turb predictions, although their longer-lag behavior differs. The agreement in autocorrelation is particularly important as it indicates that PreVAE-Turb preserves the temporal coherence of transport events.

\subsection{Efficiency}

A key advantage of surrogate models is computational speedup over direct numerical simulation. For Hasegawa-Wakatani turbulence, Clavier et al.~\cite{clavier2025generative} reported that the GAIT surrogate generates turbulent states approximately 400$\times$ faster than direct numerical integration on comparable hardware (V100 GPU). PreVAE-Turb achieves similar speedup for HW turbulence, as expected given the analogous VAE-RNN architecture.

For GENE gyrokinetic turbulence, Table~\ref{tab:efficiency} summarizes the computational cost breakdown. The one-time training cost (VAE fine-tuning plus ConvLSTM training) totals approximately 4.5 hours on an RTX A5000 GPU. Once trained, PreVAE-Turb generates trajectories rapidly: 2000 time steps in under 14 seconds. The inference time scales sublinearly with trajectory length (4000 steps in 19.4 s) due to fixed overhead from model initialization and warmup. This enables rapid generation of trajectories for statistical analysis at the trained parameter setting; parameter sweeps would require additional training or parameter conditioning.

\begin{table}[t]
\centering
\caption{Computational efficiency comparison for GENE turbulence ($128 \times 48$ grid, 4 fields). GENE evolves the full five-dimensional distribution function, whereas the surrogate predicts four two-dimensional midplane fields. Training costs are one-time (1000 epochs each); inference costs are per trajectory.}
\label{tab:efficiency}
\begin{tabular}{llcc}
\toprule
Model & Stage & Hardware & Wall Time \\
\midrule
GENE Solver & Simulation & 4$\times$ NVIDIA A100 (Perlmutter)& 235 min \\
\midrule
\multirow{3}{*}{PreVAE-Turb}
 & VAE fine-tuning & RTX A5000 & 220 min \\
 & ConvLSTM training & RTX A5000 & 53 min \\
\cmidrule{2-4}
 & Inference (2000 steps) & RTX A5000 & 13.7 s \\
 & Inference (4000 steps) & RTX A5000 & 19.4 s \\
\bottomrule
\end{tabular}
\end{table}

\section{Conclusion}
\label{sec:conclusion}

We presented PreVAE-Turb, a surrogate modeling framework for plasma turbulence that leverages pre-trained variational autoencoders from the Stable Diffusion image generation model. The framework combines the AutoencoderKL architecture for spatial compression with convolutional LSTM networks for temporal dynamics in latent space. By fine-tuning rather than training from scratch, the pre-trained VAE provides superior reconstruction fidelity compared to CNN-VAE architectures trained from scratch while significantly reducing the effort required for architecture design and hyperparameter tuning. Direct comparison on GENE gyrokinetic data demonstrates that the pre-trained approach achieves lower training loss across all reconstruction-related terms (Charbonnier, gradient, spectral) and produces surrogate models with improved validation metrics, confirming that transfer learning from natural images provides an effective foundation for turbulence compression. An ablation study shows that removing the spectral term changes POD spectra and autocorrelation structure in the evaluated rollouts, illustrating its influence on downstream statistics.

We validated the framework on two turbulence systems of increasing complexity: 2D Hasegawa-Wakatani drift-wave turbulence and gyrokinetic turbulence from the GENE code. In both cases, PreVAE-Turb accurately reproduces key statistical properties---spatial and temporal spectra, POD modes, autocorrelations, and cross-field correlations---over autoregressive rollouts extending well beyond the training horizon. For GENE, the framework handles four coupled fields simultaneously through a minimal channel adaptation, capturing multi-field spectral content, cross-correlations governing transport, and turbulent flux statistics consistent with the ground truth.

The computational efficiency of the trained surrogate enables applications that would be prohibitive with direct numerical simulation. For GENE turbulence, inference generates 2000 time steps in under 14 seconds on a single GPU, with one-time training cost of approximately 4.5 hours. This efficiency supports long-time statistical analysis at fixed parameters and motivates future extensions to parameter-dependent surrogate modeling.

The present surrogate is trained and validated for a fixed turbulence regime and a fixed set of simulation parameters; it is not intended to be used for out-of-distribution extrapolation without additional training or adaptation. Future work includes extending the framework to GENE-X simulations in edge plasma geometries, incorporating physics-informed constraints in the latent dynamics to improve long-horizon prediction fidelity, and exploring transfer learning across different plasma parameter regimes.

\FloatBarrier
\section*{Acknowledgments}
This work was supported by the U.S. Department of Energy under Contract No. DE-FG02-04ER-54742 and the Fusion Innovation Research Engine Collaborative, ``Advanced Profile Prediction for Fusion Pilot Plant Design (APP-FPP),''  under Award No. DE-SC0025853.

This research used resources of the National Energy Research Scientific Computing Center (allocation m2116), a DOE Office of Science User Facility.

\section*{Author Declarations}
\subsection*{Conflict of Interest}
The authors have no conflicts to disclose.

\subsection*{Use of AI Tools}
During the preparation of this work, the authors used Claude Fable 5 (Anthropic) and ChatGPT 6 Sol (OpenAI) for language editing and coding assistance. The authors reviewed and verified all content and take full responsibility for the content of this article.

\section*{Data Availability}
The data that support the findings of this study are available from the corresponding author upon reasonable request.

\end{document}